\documentclass[sigconf,screen]{acmart} 
\usepackage{xspace}

\usepackage[tight,footnotesize]{subfigure}

\usepackage{multirow}

\usepackage{xcolor, soul}

\usepackage[most]{tcolorbox}

\usepackage{algpseudocode}
\usepackage[algo2e,ruled,vlined,linesnumbered]{algorithm2e}

\colorlet{lightgray}{gray!30}
\sethlcolor{lightgray} 

\AtBeginDocument{%
  }

\setcopyright{acmlicensed}
\copyrightyear{2025}
\acmYear{2018}
\acmDOI{XXXXXXX.XXXXXXX}

\copyrightyear{2025}
\acmYear{2025}
\setcopyright{acmlicensed}\acmConference[FSE '25]{ Proceedings of the 33rd ACM Symposium on the Foundations of Software Engineering (FSE '25)}{June 23--27, 2025}{Trondheim, Norway}

\acmBooktitle{Companion Proceedings of the 33rd ACM Symposium on the Foundations of Software Engineering (FSE '25), June 23--27, 2025, Trondheim, Norway}

\acmISBN{978-1-4503-XXXX-X/18/06}

\newcommand*{\GS}{\texttt{GitSum}\@\xspace}

\newcommand*{\ME}{\texttt{Metagente}\@\xspace}

\newcommand*{\RM}{README.MD\@\xspace}

\newcommand*{\GH}{GitHub\@\xspace}

\newcommand*{\AL}{\texttt{Analysis Agent}\@\xspace}
\newcommand*{\CA}{\texttt{Prompt Creator Agent}\@\xspace}
\newcommand*{\EA}{\texttt{Extractor Agent}\@\xspace}
\newcommand*{\SA}{\texttt{Summarizer Agent}\@\xspace}
\newcommand*{\TA}{\texttt{Teacher Agent}\@\xspace}

\newcommand*{\LLa}{\texttt{LLaMA-2}\@\xspace}

\newcommand*{\Fo}{\texttt{GPT-4o}\@\xspace}

\newcommand*{\ie}{i.e.,\@\xspace}

\newcommand{\gpt}[1]{\texttt{\hl{\small #1}}}

\definecolor{darkgreen}{rgb}{0.0, 0.5, 0.0}
\newcommand{\TST}{\textit{TS$_{10}$}\@\xspace}
\newcommand{\TSF}{\textit{TS$_{50}$}\@\xspace}
\newcommand{\ES}{\textit{ES}\@\xspace}
\newcommand{\rqone}{\textbf{RQ$_1$}: \emph{Does the use of multi LLMs-based agents result in 
		more relevant About descriptions?}}

\newcommand{\rqtwo}{\textbf{RQ$_2$}: \emph{How does \ME perform compared to 
		\GS and \LLa?}}

\newcommand{\rqthree}{\textbf{RQ$_3$}: \emph{Which control flow between sequential and parallel fine tuning contributes to a better performance?}}

\begin{document}


\title{Amplifying the Synergy: LLMs-Based Agents for GitHub \RM Summarization}

\title{Many hands make light work: LLMs-Based Agents for GitHub \RM Summarization}

\title{Teamwork makes the dream work: LLMs-Based Agents for GitHub \RM Summarization}


\author{Duc S. H. Nguyen}
\smaller\email{duc.nsh231061m@sis.hust.edu.vn}
\small\affiliation{%
	\institution{Hanoi University of Science and Technology}
	\city{Hanoi}
	\country{Vietnam}
}

\author{Bach G. Truong}
\smaller\email{bach.tg210087@sis.hust.edu.vn}
\small\affiliation{%
	\institution{Hanoi University of Science and Technology}
	\city{Hanoi}
	\country{Vietnam}
}

\author{Phuong T. Nguyen}
\smaller\email{phuong.nguyen@univaq.it}
\small\affiliation{%
	\institution{University of L'Aquila}
	\city{67100 L'Aquila}
	\country{Italy}
}

\author{Juri Di Rocco}
\email{juri.dirocco@univaq.it}
\affiliation{%
	\institution{University of L'Aquila}	
	\city{67100 L'Aquila}	
	\country{Italy}	
}

\author{Davide Di Ruscio}
\email{davide.diruscio@univaq.it}
\affiliation{%
	\institution{University of L'Aquila}	
	\city{67100 L'Aquila}	
	\country{Italy}	
}



\begin{abstract}

The proliferation of Large Language Models (LLMs) in recent years has realized many applications in various domains. Being trained with a huge of amount of data coming from various sources, LLMs can be deployed to solve different tasks, including those in Software Engineering (SE).  
Though they have been widely adopted, the potential of using LLMs cooperatively has not been thoroughly investigated.  

In this paper, we proposed \ME as a novel approach to amplify the synergy of various LLMs. \ME is a Multi-Agent framework based on a series of LLMs to self-optimize the system through evaluation, feedback, and cooperation among specialized agents. Such a framework creates an environment where multiple agents iteratively refine and optimize prompts from various perspectives. The results of these explorations are then reviewed and aggregated by a teacher agent. To study its performance, we evaluated \ME with an SE task, \ie summarization of \RM files, and compared it with three well-established baselines, \ie \GS, \LLa, and \Fo. The results show that our proposed approach works efficiently and effectively, consuming a small amount of data for fine-tuning but still getting a high accuracy, thus substantially outperforming the baselines. The performance gain compared to \GS, the most relevant benchmark, ranges from 27.63\% to 60.43\%. More importantly, compared to using only one LLM, \ME boots up the accuracy to multiple folds. 
 
\end{abstract}

\maketitle




\section{Introduction}
\label{sec:Introduction}

%

Large Language Models (LLMs) have transformed how we approach different tasks such as natural language understanding, creative writing, and software engineering ~\cite{DBLP:journals/software/Ozkaya23b}. However, despite the individual strengths of LLMs, no single model can fully address the vast range of human language and problem domains. For example, while LLMs like GPT have excelled in natural language understanding, their performance on tasks requiring domain-specific expertise or complex multi-step reasoning remains limited \cite{He_Treude_Lo_2024}. A potential solution involves prompt-tuning, where prompts are iteratively refined to improve the performance of the pre-trained language model without modifying its 
internal design. 
Although this approach has shown promise, it faces challenges when the LLM is accessible only via an API. Furthermore, manual prompt engineering techniques, such as Chain-of-Thought (CoT) reasoning, require significant human effort to refine prompts iteratively. This labor-intensive process is prone to subjective bias and scalability issues, making it difficult to generalize across diverse tasks \cite{white2023promptpatterncatalogenhance}.

%
%

To address the inherent limitations of single LLMs, researchers have proposed multi-agent systems that enable specialized LLMs to collaborate within a shared framework~\cite{DBLP:journals/fcsc/WangMFZYZCTCLZWW24}. These systems capitalize on the unique strengths of different LLMs, where agents specialize in tasks such as code generation, debugging, or domain-specific problem-solving \cite{He_Treude_Lo_2024,DBLP:journals/corr/abs-2407-01489}. For instance, frameworks like TransAgent have demonstrated how task-specific agents can integrate seamlessly to tackle complex engineering challenges \cite{TRANSAGENT2024}. However, these systems are not without limitations. Challenges such as effective coordination, efficient communication, and the overhead of integrating multiple agents persist. Additionally, designing robust frameworks for agent interaction and feedback loops requires considerable engineering and computational resources. Despite these constraints, the collaborative potential of multi-agent systems offers an interesting alternative to relying on single model's capabilities~\cite{10.1145/3691620.3695291}.


To illustrate a practical application of multi-agent LLM systems, we propose a novel framework targeting the problem of summarizing SE artifacts. This approach leverages the capabilities of specialized LLM agents to optimize task performance. The work supports a call for fundamentally new research directions with an initial evaluation on a real issue in SE, thus having 
the following contributions: 

\smallskip
\noindent
$\triangleright$ \textbf{Solution}. We develop \ME, an LLMs-based agents approach to perform the summarization of SE artifacts. A reciprocal teacher-student architecture is built with 
two components, \ie the master module to perform the main task, 
and the optimization module 
to refine and enhance the master module.
To the best of our knowledge, our work is the first one ever to study the applicability of LLMs-based agents in this domain. 

\smallskip
\noindent
$\triangleright$ \textbf{Evaluation}.
Even though our work 
explores a new direction, being still in early stages of research, we supported by initial evidence with a concrete SE task. In particular, we used the pipeline to perform the summarization of \GH \RM files. This task has been specifically chosen due to the following reasons: \emph{(i)} Many \GH repositories do not have an About description~\cite{10.1145/3593434.3593448}, posing difficulties to users when getting acquainted with their content; and \emph{(ii)} The diversity of text, mark down content, and source code in \RM files makes it difficult to produce a good summary. 

\smallskip
\noindent
$\triangleright$ \textbf{Comparison}. An empirical evaluation has been conducted to compare 
\ME with \GS~\cite{10.1145/3593434.3593448} and \LLa, two state-of-the-art baselines. 

\smallskip
\noindent
$\triangleright$ \textbf{Open Science}. A replication package including the dataset and source code of \ME has been published to foster future research~\cite{MetagenteArtifacts}.


\section{Related Work}
\label{sec:RelatedWork}

\subsection{LLMs-based Multi-Agent Systems}

Integrating multi-agent systems and LLMs has facilitated the development of adaptive systems that enhance collaboration and productivity in SE. Several frameworks have emerged to address specific challenges in this domain, 
focusing on collaboration, modularity, and optimization.

OPRO~\cite{DBLP:conf/iclr/Yang0LLLZC24}  employs different LLMs as optimizers, leveraging natural language prompts to generate and refine solutions iteratively. Similarly,  APE~\cite{DBLP:conf/iclr/ZhouMHPPCB23} generates candidate instructions and iteratively refines them using semantic similarity and evaluation metrics. 
Other frameworks, such as Camel~\cite{10.5555/3666122.3668386} and MetaGPT~\cite{hong2024metagpt}, emphasize modularity and structured collaboration among agents. Camel introduces a role-playing framework that guides chat agents using inception prompting, aligning their actions with human intentions. This 
enables instruction-following cooperation and generates conversational data that advances the understanding of multi-agent behaviors. MetaGPT, on the other hand, incorporates human workflows as a meta-programming approach to address challenges like 
hallucination. 

Frameworks like AutoGen~\cite{wu2024autogen} and LangChain\footnote{\url{https://www.langchain.com/}} focus on leveraging LLMs for modular and conversation-driven application development. AutoGen facilitates the creation of LLM-based applications through conversable agents that can autonomously engage in multi-turn conversations, incorporate human feedback, and combine modular capabilities. Its conversation programming paradigm simplifies the definition of agent roles and interaction behaviors across a variety of domains, including coding and operations research. LangChain complements this effort by providing components and customizable pipelines for integrating external data sources and interacting with other applications. Its modular abstractions and chains streamline the development of applications~\cite{langchain,10242497,soygazi2023analysis,10.1145/3691620.3695291}. 

\subsection{LLMs for summarization tasks}
 
The SE 
community has increasingly explored the application of summarization techniques across a variety of tasks \cite{MastropaoloCPB24}. 
Integrating LLMs and pre-trained model into text summarization has garnered significant attention due to their potential to streamline various domain-specific tasks. 
A recent study~\cite{10.1007/978-3-031-78107-0_10} compared five pre-trained models across three different natural language processing (NLP) tasks: text classification, summarization, and generation. The results of the text summarization task revealed that GPT, BART, and LLaMA achieved the highest accuracy among others. 

Prompt engineering has been explored in various domains as well. Bajaj and Borhan~\cite{bajaj_effect_2024} compared zero-shot, few-shot, and role-based prompting techniques for summarizing diverse articles. Their findings emphasized the versatility of simple zero-shot prompts, which consistently outperformed other methods. Similarly, Oliveira and Lins~\cite{oliveira_assessing_2024} evaluated both abstractive and extractive summarization methods, highlighting the superior performance of Pegasus for news summarization tasks.

%

%
Doan et al.~\cite{10.1145/3593434.3593448} introduced \GS, a novel approach to summarizing \RM content. \GS is built upon 
BART and T5 
to recommend descriptions for repositories that lack such metadata. Similarly, FILLER
~\cite{10.1145/3674805.3686682} is a solution for generating titles for Stack Overflow posts by leveraging a fine-tuned language model equipped with self-improvement mechanisms and post-ranking capabilities. FILLER adopts a multi-task learning strategy, fine-tuning the CodeT5 model on a dataset of Stack Overflow posts while simultaneously training across multiple programming languages \cite{10.1145/3674805.3686682}.


\section{Proposed Approach}
\label{sec:ProposedApproach}

\begin{figure}[t!]
	\centering
	\includegraphics[width=1.0\columnwidth]{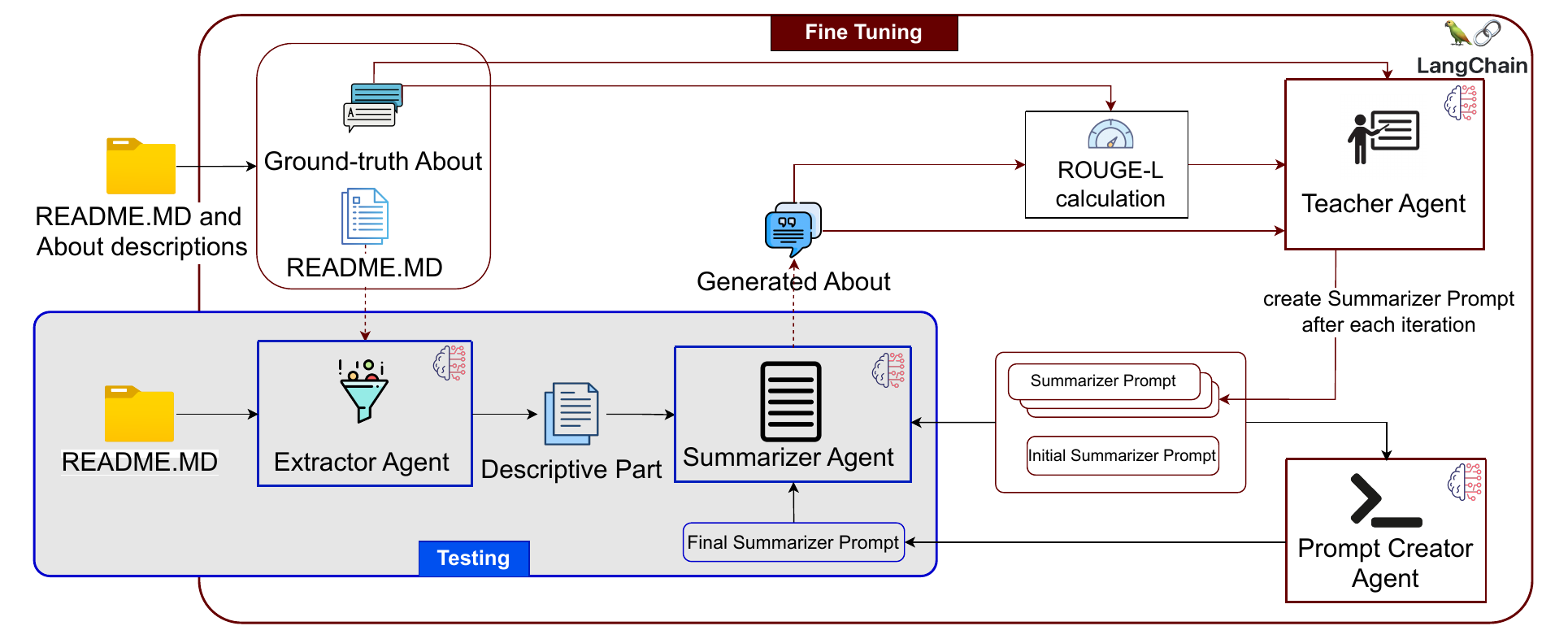}
	\caption{The proposed \ME pipeline.}
	\label{fig:Architecture}
	\vspace{-.2cm}
\end{figure}


Fig.~\ref{fig:Architecture} depicts the proposed framework to 
create an 
environment where multiple LLMs-based agents iteratively refine and optimize the outcome. 
It is 
a teacher-student architecture 
with 
2 components: the main module, which generates 
``About'' content from \RM text, and the optimization module, which participates only in the training phase to refine and enhance the main module. %

\subsection{LLMs-based Agents}



The pipeline in Fig.~\ref{fig:Architecture} consists of 4 agents, \ie \EA, \SA, \TA, and \CA. Throughout the pipeline, prompts are used 
to instruct the 
agents, and a scoring mechanism based on the ROUGE metrics is utilized to guide the LLMs in identifying the most suitable actions. \emph{Due to the strict space limit, we cannot show the prompts 
here, but upload them to 
the online appendix for the sake of references}~\cite{MetagenteArtifacts}. 




\smallskip
\noindent
$\triangleright$ \EA. Being steered by prompts, this agent removes irrelevant and noisy details from input \RM files, focusing only on content that introduces or describes the repository. 
A \RM 
might contain many sections such as introduction, description, installation, contributing, license, and \EA needs to filter out all sections 
not relevant to the 
description of the repository 
to yield a short and concise descriptive text. 
During optimization, only this descriptive text is used instead of the entire \RM file, so as to significantly reduce time and computational resources.

\noindent
$\triangleright$ \SA. 
%
It begins with a predefined initial prompt and uses it to summarize the extracted text from a \RM file. Through an iterative process, during each iteration the agent takes as input an updated summarizer prompt generated by the \TA. The goal is to produce a concise About description that captures the core concept or purpose of the repository, focusing on key terms, features, and specific context without including explanations or extraneous details. The final output should be a short and precise phrase that enhances clarity and relevance while reflecting the repository's essential idea.

\noindent
$\triangleright$ \TA. 
By reviewing the following four inputs: \emph{(1)} current \SA prompt, \emph{(2)} generated About, \emph{(3)} ground-truth About, and \emph{(4)} ROUGE-L scores, 
this agent 
optimizes \SA’s prompt for each training sample. A complex prompt with various steps guides \SA in 
analyzing the input features, and 
improving the current prompt of \SA. 


\noindent
$\triangleright$ \CA. Accepting as input a set of prompts, the agent analyzes and identifies common parts to produce the final prompt. 
It extracts specific details or conditional key points from the input prompts to be included in the final prompt. 
Being derived from these seed data instances, the final prompt serves as an overall guide for \SA’s inference task. 
It provides high-level instructions while offering specific guidance on key details tailored to the nuances of the training data.



For \EA and \SA, we opted for OpenAI’s \gpt{GPT-4o-mini} 
as the LLM engine. 
For 
\TA and \CA, we leveraged the more advanced \gpt{GPT-4o} model. This hierarchical deployment ensures that the larger and more advanced LLMs guide the smaller models, optimizing their performance without incurring excessive computational costs during inference. 
This helps the final system maintain a balance of high efficiency, cost-effectiveness, and optimal performance.

\subsection{Orchestration of Agents}


\ME is a cooperative pipeline 
where LLMs-based agents work together towards a shared objective by means of 
three main phases, \ie 
\textit{Communications}, \textit{Self Improvement}, and \textit{Prompt Generation} as follows.


\smallskip
\noindent
$\triangleright$ \textbf{\texttt{Communications}}. An agent works by interacting with the environment and other agents. 
In our pipeline, besides LangChain, 
that has been adopted as the communication backbone, 
we also employed a structured output mechanism to guide the LLMs to produce information that is easily digestible as input for the subsequent agents. This is to ensure consistency and seamlessness in exchanging messages among the agents. 




\smallskip
\noindent
$\triangleright$ \textbf{\texttt{Self Improvement}}. 
At the beginning of each iteration, the extracted text is fed as input to \SA. 
Based on the current prompt 
for that iteration, \SA generates a short description as output. To evaluate the generated About, we use the ROUGE metrics, which have been widely adopted for text summarization evaluation~\cite{10.1145/3593434.3593448,10.1145/3324884.3416538,10.1145/3401026}. Specifically, we focus on the ROUGE-L score to facilitate the comparison of results across iterations. The current \SA prompt, generated About, ground-truth About, and ROUGE-L scores are used as input for \TA. Being guided with dedicated prompts, the agent compares the generated About with the ground-truth About to find out the key differences, and propose the necessary improvements to \SA’s prompt. 

\TA generates a new prompt, which is 
used by \SA in the next iteration. During the optimization process, we impose a limit on the number of iterations. The termination is met when: \emph{(i)} The ROUGE-L score of the current generated About reaches a predefined threshold; \emph{(ii)} or the maximum number of iterations is exceeded.
After experimenting with multiple configurations, we set the following hyperparameters: \emph{(i)} ROUGE-L threshold = $0.7$; and maximum number of iterations = $15$. 

At the end of the optimization, 
we obtain a final 
prompt for each training data instance. Instances failing to meet the ROUGE-L threshold within the allowed iterations are discarded. The remaining successful data instances are referred to as seed training data, as they best represent and generalize the dataset in terms of converting \RM text into an About description.

\smallskip
\noindent
$\triangleright$ \textbf{\texttt{Prompt Generation}}. 
The optimization flow on a selected dataset results in a set of distinct prompts for \SA, each of them is optimized to deliver strong performance on a specific data instance. This set of prompts is then passed as the sole input for \CA, which 
extracts common instructions shared across all candidate prompts, while identifying and integrating key conditional points. These points guide \SA in adapting its structure and writing style to suit the varying contexts and characteristics of different \RM, ensuring consistent and high-quality About content generation under diverse scenarios.

\section{Proof of Concept}
\label{sec:Evaluation}

\subsection{Research questions}

We perform various 
experiments to answer the following research questions.

\smallskip
\noindent
$\triangleright$ \rqone~This research question compares \ME with a system that has only one \Fo summarization engine. We study if the combination of a series of LLMs is really needed, considering the fact that a single LLM agent might possibly 
be sufficient to get a good recommendation performance. 
This is important in practice as a simple yet effective approach is preferable in the era of Generative AI. 

\smallskip
\noindent
$\triangleright$ \rqtwo~
We compare \ME with \GS~\cite{10.1145/3593434.3593448}--a state-of-the-art tool in summarizing \RM files, and 
\LLa--a large language model that has been widely applied in various SE tasks~\cite{10.1145/3695988}, and 
summarization~\cite{DBLP:conf/icpr/PathakR24,10740129}. 

\begin{figure*}[t!]
	\centering
	\begin{tabular}{c c c}	
		\subfigure[ROUGE-1: Average 0.409, 0.162, 0.282, \textbf{0.522} ({\color{darkgreen} $\uparrow$27.63\%, $\uparrow$222.22\%, $\uparrow$85.11\%}) ]{\label{fig:RQ1-ROUGE1-50samples}
			\includegraphics[width=0.22\linewidth]{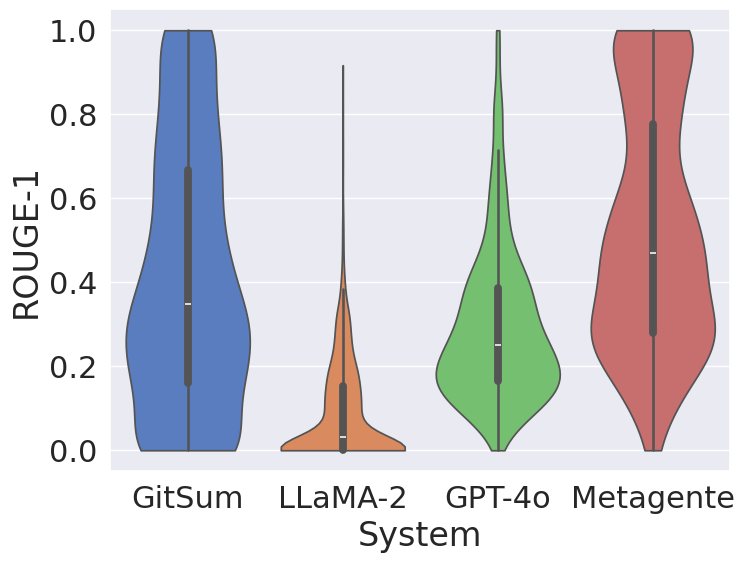}}	&
		\subfigure[ROUGE-2: Average 0.272, 0.100, 0.152, \textbf{0.363} ({\color{darkgreen} $\uparrow$33.46\%, $\uparrow$263.00\%, $\uparrow$138.82\%}) ]{\label{fig:RQ1-ROUGE2-50samples}\includegraphics[width=0.22\linewidth]{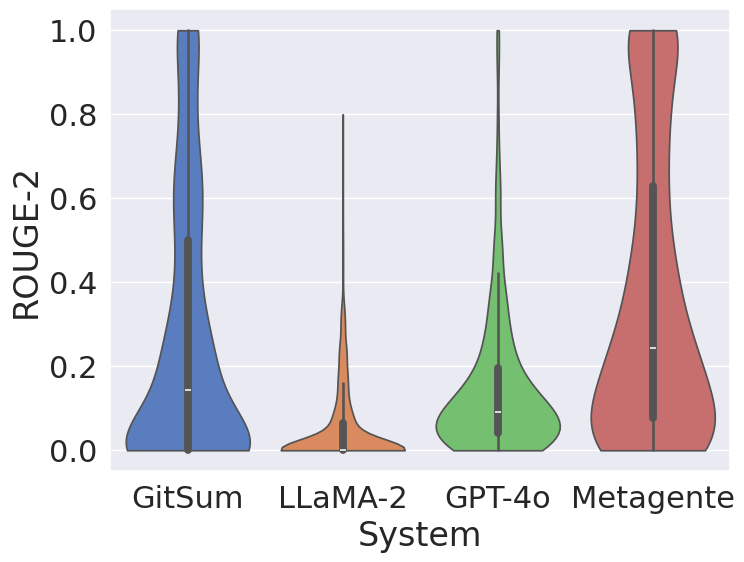}} &
		\subfigure[ROUGE-L: Average 0.387, 0.146, 0.250, \textbf{0.486} ({\color{darkgreen} $\uparrow$25.58\%, $\uparrow$232.88\%, $\uparrow$94.40\%}) ]{\label{fig:RQ1-ROUGEL-50samples}\includegraphics[width=0.22\linewidth]{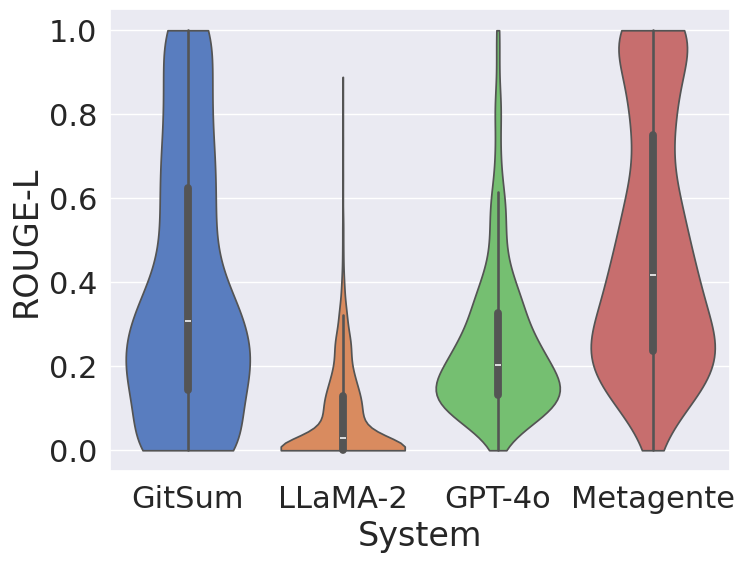}} 
	\end{tabular} 
	\vspace{-.2cm}
	\caption{Comparison of \GS, \LLa, \Fo, and \ME: \TSF is used for training and fine tuning.} 	
	\label{fig:RQ1-50samples}
		\vspace{-.2cm}
\end{figure*}

\begin{figure*}[t!]
	\centering
	\begin{tabular}{c c c}	
		\subfigure[ROUGE-1: Average 0.360, 0.088, 0.297, \textbf{0.536} ({\color{darkgreen} $\uparrow$48.89\%, $\uparrow$509.09\%, $\uparrow$80.47\%}) ]{\label{fig:RQ1-ROUGE1-10samples}
			\includegraphics[width=0.22\linewidth]{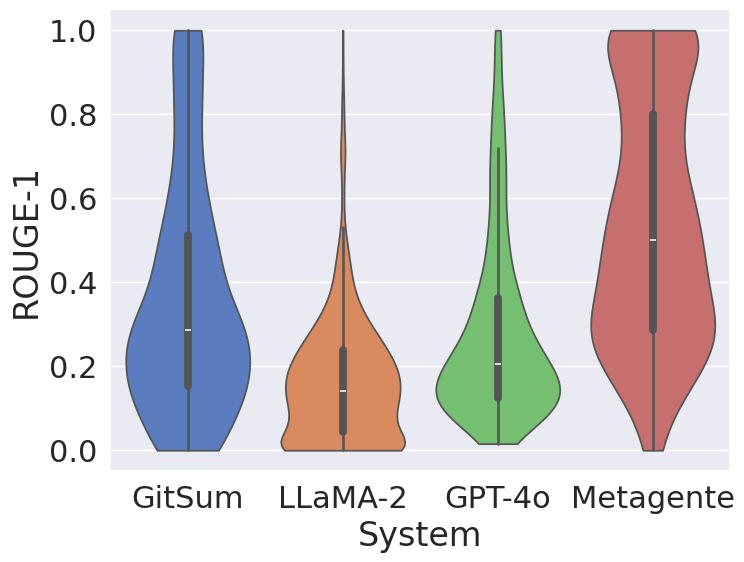}}	&
		\subfigure[ROUGE-2: Average 0.235, 0.049, 0.151, \textbf{0.377} ({\color{darkgreen} $\uparrow$60.43\%, $\uparrow$669.39\%, $\uparrow$149.67\%}) ]{\label{fig:RQ1-ROUGE2-10samples}\includegraphics[width=0.22\linewidth]{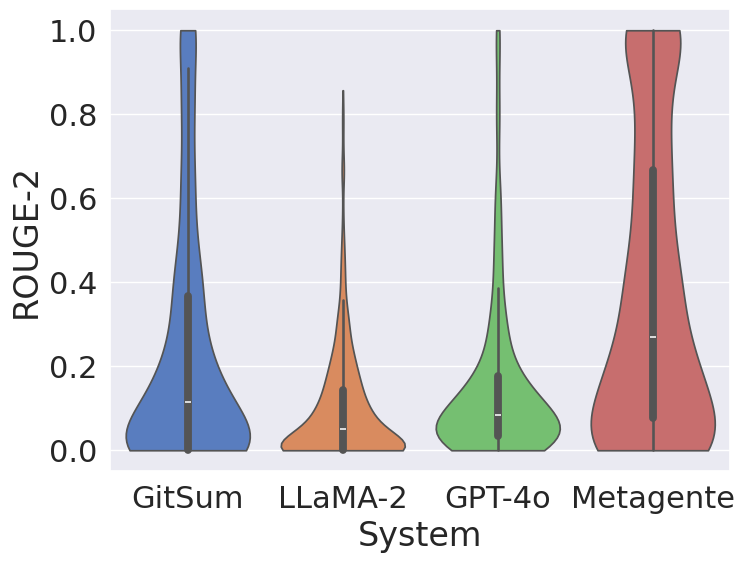}} &
		\subfigure[ROUGE-L: Average 0.334, 0.079, 0.256, \textbf{0.503} ({\color{darkgreen} $\uparrow$50.60\%, $\uparrow$536.71\%, $\uparrow$96.48\%}) ]{\label{fig:RQ1-ROUGEL-10samples}\includegraphics[width=0.22\linewidth]{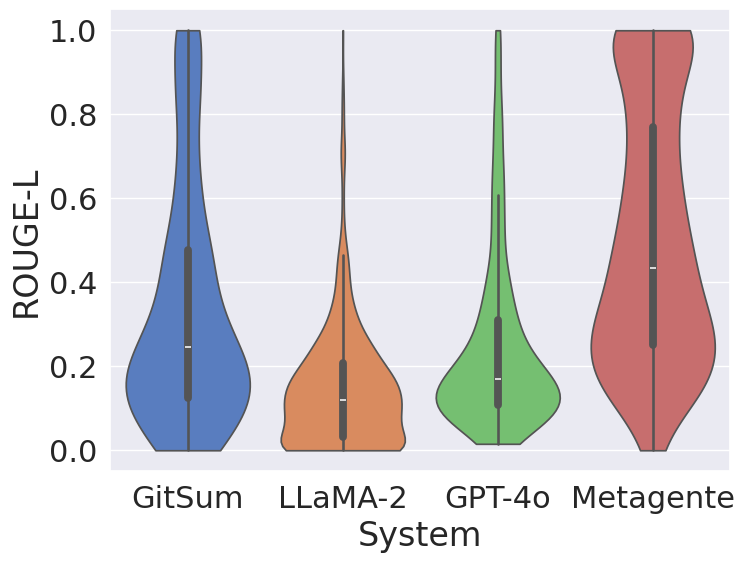}}
	\end{tabular} 
	\vspace{-.2cm}
	\caption{Comparison of \GS, \LLa, \Fo, and \ME: \TST is used for training and fine tuning.} 	
	\label{fig:RQ1-10samples}
		\vspace{-.2cm}
\end{figure*}

\subsection{Dataset and Metrics}

We adopted an existing dataset for \RM-related tasks~\cite{10.1145/3593434.3593448}, and extended it by incorporating a diverse range of data sources to enhance its comprehensiveness and applicability. First, we augmented the initial dataset with \GH repositories categorized under \gpt{awesome-lists} and \gpt{documentation-related} topics\footnote{\url{https://github.com/topics}} that align with the document repositories category~\cite{zanartu_automatically_2022}. Then, we 
enriched the dataset by including curated repositories containing popular Python projects~\cite{2021-03-23-popular-3k-python_dataset_2023}, Jupyter notebooks for data analysis~\cite{10.1145/3173574.3173606,borges_understanding_2016}. 
This approach is motivated by the observation that popular repositories often feature well-maintained and detailed \RM files~\cite{wang_study_2022}.
As a result, we compiled 6,933 unique repositories containing at least a \RM file. 
Through a manual inspection, we noticed that by several repositories, the About descriptions do not match with what was written by the \RM files. This happens because developers changed the \RM files, but then forgot to update the corresponding About. Thus, we filtered out those 
and eventually 
obtained 
925 samples, from which 2 training sets were randomly selected with 10 and 50 samples to yield  \TST and \TSF, and a testing set of 865 samples named as \ES. 

To evaluate the recommendations, we use the ROUGE metrics, \ie ROUGE-1, ROUGE-2, ROUGE-L, which have been widely used in text summarization \cite{Chen2020,itiger2022}. Due to space limit, we cannot recall them here, interested readers are kindly referred to the work of Lin~\cite{rouge2004} for greater details.

\subsection{Settings}

During the experiments, we observed that \EA adapted effectively to various \RM texts without encountering significant challenges. Thus, 
the prompt for \EA was pre-optimized and remained fixed throughout the entire optimization. 
We compared 
\ME with three baselines: \GS (based on BART), fine-tuned \LLa, and \Fo. These methods were identified as the most promising ones in a recent study~\cite{10.1007/978-3-031-78107-0_10}.
For the first experiment, we used \TSF with 50 \RM files for training \ME and the selected baselines. In the second experiment, we decreased the number of samples 
to 10, \ie the \TST dataset. This aims to study whether the tools are capable of generating About descriptions when an extremely small amount of data is available for training. 
This is useful in practice, as curating a suitable dataset for training is both time consuming and prone to error, and a model that can produce relevant About descriptions given a limited amount of training data is preferable.

\section{Results and Discussion}
\label{sec:Results}

Fig.~\ref{fig:RQ1-50samples} and Fig.~\ref{fig:RQ1-10samples} report the ROUGE scores using violin boxplots. By each subfigure, we compute the average scores of \GS, \LLa, \Fo, and \ME. 
In brackets we depict the increase in percent (a green up arrow {\color{darkgreen} $\uparrow$}) 
of the average score of \ME compared to those of the 
baselines. 

\subsection{Result Analysis}



\smallskip
\noindent
$\triangleright$ \textbf{\rqone}

For the comparison between a single \Fo 
and \ME, we consider the third and fourth boxplots in Fig.~\ref{fig:RQ1-50samples} and Fig.~\ref{fig:RQ1-10samples}, corresponding to the use of \TSF and \TST for fine tuning (training), respectively and \ES for testing. Overall, it is evident that \ME obtains a better accuracy compared to that of \Fo in terms of ROUGE-1, ROUGE-2, and ROUGE-L. 
As shown in Fig.~\ref{fig:RQ1-ROUGE1-50samples}, by most of the testing instances, \Fo gets a ROUGE-1 score 
lower than $0.3$. Meanwhile, \ME performs better as the density of the violin boxplots is concentrated on the 0.5 to 1.0 range, corresponding to superior ROUGE-1 scores by the majority of the testing samples. Looking at the average score, we see that \Fo and \ME get 0.282 and 0.522, respectively, resulting in 
an increase of $85.11\%$ by \ME compared to \Fo. Such a gain is greater by the ROUGE-2 scores in Fig.~\ref{fig:RQ1-ROUGE2-50samples}, which shows that \ME gets $0.363$ as ROUGE-2 score, being greater than $0.152$, the corresponding score obtained by \Fo, yielding an improvement of $138.82\%$. The same trend is seen with the ROUGE-L scores (Fig.~\ref{fig:RQ1-ROUGEL-50samples}), \ie \ME performs better than \Fo.

When fewer samples are used 
for training, \ie \TST includes only 10 \RM files and About descriptions, as shown in Fig.~\ref{fig:RQ1-10samples}, the gain 
by \ME compared to \Fo is much greater. In particular, the increase is $80.47\%$, $149.67\%$, and $96.48\%$ by ROUGE-1, ROUGE-2, and ROUGE-L, respectively. Such a difference implies that \ME is effective, even when there is a limited amount of data for fine tuning.


We ran Wilcoxon rank tests on every pair of ROUGE scores of \ME and \Fo for the whole testing set with 865 samples, and got the following p-values: p=8.62e-89 (ROUGE-1), p=3.40e-72 (ROUGE-2), p=4.91e-91 (ROUGE-L). %
The rank tests confirm that \emph{the performance difference obtained by \ME in relation to \Fo is statistically significant}.


\vspace{.2cm}
\noindent\fbox{\begin{minipage}{0.98\columnwidth}
		\paragraph{\textbf{Answer to RQ$_1$:}} Compared to a single LLM agent, the combination of multi LLMs-based agents is clearly advantageous, 
		as it brings a lot 
		more precise About descriptions 
		with respect to all the ROUGE scores. 
\end{minipage}}

\vspace{.3cm}
\noindent
$\triangleright$ \textbf{\rqtwo}


We refer to Fig.~\ref{fig:RQ1-50samples} and Fig.~\ref{fig:RQ1-10samples} again for the comparison between \ME and the two baselines, \ie \GS and \LLa on the \TSF and \TST datasets. Concerning ROUGE-1, as shown in Fig.~\ref{fig:RQ1-ROUGE1-50samples}, \GS is much better with respect to \LLa as most of its scores are scattered between 0.2 and 0.5; it also obtains ROUGE-1 score of 1.0 by different testing samples, while the accuracy of \LLa is much lower, with no score being seen at the 1.0 level. \ME outperforms both baselines since it has more scores on the range from 0.6 to 1.0. The density of the 1.0 level by \ME is also larger than that of \GS and \Fo. On the average score, \ME gets 0.522, which is better than 0.409 and 0.162--the corresponding values achieved by \GS and \LLa, resulting in a gain of
$27.63\%$ and $222.22\%$, respectively. The difference in performance of \ME compared to the baselines 
is more evident with ROUGE-2 in Fig.~\ref{fig:RQ1-ROUGE2-50samples}, \ie $33.46\%$ and $263.00\%$; 
and ROUGE-L in Fig.~\ref{fig:RQ1-ROUGEL-50samples}, \ie $25.58\%$ and $232.88\%$.

Similarly, the results for the \TST dataset in Fig.~\ref{fig:RQ1-10samples} show that 
\ME outweighs the baselines by all the three 
metrics. Especially, with the average ROUGE-2 score, \ME yields an increase of $669.39\%$ compared to \LLa. Wilcoxon rank tests reveal that the performance gain is always statistically significant as all the p-values are much smaller than 5e-2.\footnote{Due to space limit, we publish the statistical test results in the replication package~\cite{MetagenteArtifacts}.}




\vspace{.2cm}
\noindent\fbox{\begin{minipage}{0.98\columnwidth}
		\paragraph{\textbf{Answer to RQ$_2$:}} \ME outperforms both \GS and \LLa by all the three metrics, \ie ROUGE-1, ROUGE-2, and ROUGE-L. 
		Our tool is robust even when there is a small amount of data for fine tuning.
\end{minipage}}

\vspace{.2cm}
\noindent
$\triangleright$ \textbf{Timing performance}.
An important factor 
is the timing efficiency of 
\ME for fine tuning and testing. 
While the pipeline was implemented by us to orchestra the agents, and run on a normal laptop, all the computations are performed on OpenAI's servers, since \gpt{GPT-4o-mini} and \gpt{GPT-4o} are utilized as the LLM engines. 
We recorded the time and got the following information: With the \TSF dataset, \ME takes 3 minutes for fine tuning and 8 minutes for testing on \ES (865 samples). 

\subsection{Discussion}

\smallskip
\noindent
$\triangleright$ \textbf{Applicability.} The 
evaluation 
demonstrates that the use of a single \Fo agent does not suffice to produce 
precise About descriptions. Thus, 
our proposed pipeline is meaningful since it combines the strength of different LLMs-based agents to perform the 
task. The agents reciprocally enforce each other by means of prompts, being able to self-optimize through evaluation and feedback. 
In other words, teamwork allows LLMs to augment the synergies among them, thus achieving a superior performance. 
\ME needs a few samples for fine tuning, and this is practical in real-world usages, as curating a proper dataset is time consuming and error prone. 

\smallskip
\noindent
$\triangleright$ \textbf{Limitations.} While \ME obtains a promising performance for the dedicated task, there are different aspects to be considered. We suppose that the prompts used to guide the agents could be further optimized to improve the overall performance. In this paper, 4 agents were used for the summarization, depending on the tasks, we may extend the architecture to have more tailored LLMs-based agents. 


\smallskip
\noindent
$\triangleright$ \textbf{Threats to Validity.} \emph{(i) Internal validity}: We compared \ME with \GS using the original implementation of \GS. 
For the comparison 
with the baselines, 
we used the same set of data for fine tuning (training), and testing; \emph{(ii) External validity}: The findings are valid for the dataset used in this paper. Repositories collected from \GH are heterogeneous, and thus requiring additional prompts to preprocess and clean the data.

%

\section{Conclusion and Future work}
\label{sec:Conclusion}


In this paper, we conceptualized \ME as a practical approach to tackle the issue of summarization for \GH \RM files, leveraging 
LLMs-based agents. An empirical evaluation conducted on \ME using a dataset collected from \GH showed that our proposed approach augments the strengths of different agents, and thus outperforming a single agent as well as two state-of-the-art baselines. For future work, we plan to study the applicability of the framework in other domains in Software Engineering, such as code completion, or code review. 

\begin{acks} 
	
	This paper was partially supported by the MOSAICO project (Management, Orchestration and Supervision of AI-agent COmmunities for reliable AI in software engineering) that has received funding from the European Union under the Horizon Research and Innovation Action (Grant Agreement No. 101189664). The work has been partially supported by the EMELIOT national research project, which has been funded by the MUR under the PRIN 2020 program (Contract 2020W3A5FY). It has been also partially supported by the European Union--NextGenerationEU through the Italian Ministry of University and Research, Projects PRIN 2022 PNRR \emph{``FRINGE: context-aware FaiRness engineerING in complex software systEms''} grant n. P2022553SL. We acknowledge the Italian ``PRIN 2022'' project TRex-SE: \emph{``Trustworthy Recommenders for Software Engineers,''} grant n. 2022LKJWHC. 
\end{acks}

\bibliographystyle{ACM-Reference-Format}
\bibliography{main}


\end{document}